\documentstyle[epsfig]{aipproc}

\def\fbi {{\rm fb}^{-1}}

\newcommand{\lsim}{\mathrel{\lower4pt\hbox{$\sim$}}
\hskip-14.5pt\raise1.6pt\hbox{$<$}\;}

\newcommand{\gsim}{\mathrel{\lower4pt\hbox{$\sim$}}
\hskip-13.5pt\raise1.6pt\hbox{$>$}\;}

\newcommand{\cP}{{\cal P}}
\newcommand{\cL}{{\cal L}}

\def\ti    {\tilde}

\def\st    {{\ti t}}

\def\stau  {{\ti\tau}}
\def\snu   {{\ti\nu}}
\def\sl   {{\ti\ell}}
\def\ch    {\ti \chi}
\def\nt    {\ti \chi^0}

\def\cth   {\cos\theta}

\def\smu   {{\ti\mu}}
\def\smul  {{\ti\mu}_L}
\def\smur  {{\ti\mu}_R}
\def\se    {{\ti e}}
\def\sel   {{\ti e}_L}
\def\ser   {{\ti e}_R}

\newcommand{\hdick}{\noalign{\hrule height1.4pt}}
\newcommand{\MeV}{\mathrm{MeV}}
\newcommand{\GeV}{\mathrm{GeV}}
\newcommand{\TeV}{\mathrm{TeV}}

\newcommand{\dmchi}{\Delta m_{\tilde\chi_1}}

\begin{document}

\title{Supersymmetry and Superpartners}

\author{Hans-Ulrich Martyn}
\address{I. Physikalisches Institut der RWTH, D--52056 Aachen, Germany}

\maketitle

\begin{abstract}
  A brief summary is given of studies on supersymmetry and the spectrum
  of superparticles presented 
  at the Linear Collider Workshop 2000.
\end{abstract}

\section{Introduction}  
Although the Standard Model (SM) is extremely successful, many physicist
believe that new physics will show up at the TeV scale. 
Supersymmetry (SUSY) is considered as the most attractive extension of the 
Standard Model, in particular in its variant of the 
Minimal Supersymmetric Standard Model (MSSM).
The open question of how supersymmetry is broken and how this breaking is 
communicated to the particles has been discussed by Godbole~\cite{godbole}.
Frequently used schemes are the minimal supergravity (mSUGRA) model, 
gauge mediated (GMSB), gaugino mediated ($\chi$MSB)
and anomaly mediated (AMSB) supersymmetry breaking models, which all lead to
quite different phenomenological implications.
In many scenarios at least some particles of the supersymmetric spectrum 
--- especially the non-coloured charginos, neutralinos and sleptons --- 
are expected to be light enough to be accessible at one of the proposed 
$e^+e^-$ Linear Colliders operating at $\sqrt{s} = 0.5 - 1~\TeV$.

Since the last workshop of this series, LCWS~99~\cite{lcws99}, some ongoing
studies on supersymmetry and the superpartners of the SM particles
have been completed and many new ideas have been addressed.

A major concern is, what collider energy is required?
Cosmological constraints have been discussed by Feng~\cite{feng}.
If the lightest neutralino $\nt_1$ is taken as cold dark matter candidate
in mSUGRA models, cosmology does not provide useful bounds on 
superpartner masses.
A possibly large scalar mass $m_0 \gtrsim 1~\TeV$ would lead to
heavy sleptons without affecting the low mass gaugino sector.
But if SUSY should be observable at $500~\GeV$, then some signature of
supersymmetry should already show up before the LHC, which would be very
exciting.

Arguments of naturalness and fine tuning were examined by 
Anderson~\cite{anderson}. He comes to the conclusion that if supersymmetry
is relevant to the weak scale, it should provide a multitude of sparticles,
but probably not the complete spectrum, kinematically accessible
at a $1~\TeV$ Linear Collider.

Murayama~\cite{murayama} pointed out that in order to establish SUSY and
to remove ambiguous interpretations by alternative theories, 
as many observables of the sparticle spectrum as possible have to be measured 
and proven to be consistent with supersymmetry. 
Each SM particle has to have a superpartner with 
a spin differing by $1/2$, the same gauge quantum numbers and
identical couplings, for example $g_{e\nu W} = g_{e \snu \ti W}$.
This will be a long-term programme and the required precision can only be 
achieved at an $e^+e^-$ Linear Collider.

\section{Polarisation}
A very important tool to study supersymmetry will be the use of
highly polarised beams, as discussed by Moortgat-Pick~\cite{moortgat}.
Performances of ${\cal P}_{e^-}=0.8$ and ${\cal P}_{e^+}=0.6$ 
appear feasible.
A proper choice of polarisations and centre of mass energy helps disentangle
the particle spectrum by enhancing specific reactions and 
suppressing unwanted background.
Electron polarisation is absolutely essential to determine the weak quantum 
numbers, couplings and mixings,
{\em e.g.} to associate the chiral couplings of the right-handed and 
left-handed fermions to their $R$, $L$ superpartners.
Positron polarisation offers additional important advantages 
by selecting respectively enriching 
initial states of a definite spin:
(i) it provides an improved separation of sparticle production and decay 
topologies and thus a higher precision on model parameters by exploiting all 
combinations of polarisations;
(ii) it increases the event rate (factor 1.5 or more) 
resulting in a higher sensitivity to rare decays and subtle effects;
and (iii) it strongly supports the discovery of new physics, 
{\em e.g.} the exchange of spin 0 particles.
Note that all types of helicity conserving processes 
(SM and alternative theories)
profit in increased rates by having both $e^\pm$ beams polarised.
A few examples should illustrate the impact of positron polarisation.

$\rhd$
An interesting case is associated selectron production 
$e^-e^+ \to \ser \sel$ via $t$ channel $\nt$ exchange.
Using polarised beams the charge of the observed lepton can be directly 
related to the $L,\, R$ quantum number of the produced selectron, 
$e^-_{L,\,R} \to \se^-_{L,\,R}$ and $e^+_{L,\,R} \to \se^+_{L,\,R}$
at the corresponding vertex.
This elegant separation of selectron species and their decay spectra has been 
proposed for polarised electrons in the talk by Dima~\cite{dima}. 
But obviously the method
can be efficiently improved if the $e^+$ beam is polarised as well.

$\rhd$
Stop quarks are expected to have large mixings between $\ti t_L$ and 
$\ti t_R$,  {\em e.g.} the lighter state being
$\ti t_1 = \ti t_L \cth_{\ti t} + \ti t_R\sin\theta_{\ti t}$.
The polarised cross sections $e_L^+e_R^-\to\st_1 \st_1$
and $e_R^+e_L^-\to\st_1 \st_1$
have a characteristic dependence on the mass and stop mixing angle 
$\theta_{\st}$.
A simulation of stop decays $\st_1 \to c\,\nt_1, \ b\,\ch^+_1$ shows
that the sensitivity to both parameters can be
increased by more than $20\,\%$ when choosing the maximal
$e^-$ {\it and} $e^+$ polarisation, 
see contribution of Sopczak~\cite{sopczak}.

$\rhd$
Gaugino production $e^+e^-\to \ch^+_i\ch^-_j, \ \nt_i\nt_j$ exhibit
a very pronounced polarisation dependence. 
With polarised positrons the error on SUSY parameters 
and the masses of exchanged sleptons
can be substantially reduced. 
In particular the reach and separation towards extended models
with low cross sections, such as NMSSM and E6, 
can be largely extended~\cite{moortgat}.

$\rhd$
Very spectacular signatures would arise from rare processes with `wrong' 
helicities, which are absent in the Standard Model. 
Such reactions may occur through spin 0 sparticle exchange in $R_p$ violating 
SUSY models.
Examples are resonant or contact interaction type fermion pair production
$e^+e^- \to \tilde\nu \to \ell \bar\ell$
and single neutralino or chargino production
$e^+e^-  \to \tilde\nu \to \nu\nt, \ \ell^\pm \chi^{\mp}$
mediated through $s$-channel sneutrino exchange.

\section{Sleptons}
Scalar leptons are the superpartners of the right-handed and left-handed 
leptons. They are produced in pairs 
\begin{eqnarray*}
   e^+e^- & \to &  \ser  \ser, \ \sel \sel, \ \ser \sel,\  \snu_e\bar{\snu}_e 
     \\
   e^+e^- & \to &  \smur  \smur, \ \smul \smul, \ \snu_\mu\bar{\snu}_\mu
   \\
   e^+e^- & \to &  \stau_1  \stau_1, \ \stau_2 \stau_2, \ \stau_1 \stau_2, \
                   \snu_\tau\bar{\snu}_\tau
\end{eqnarray*}
via $s$-channel $\gamma / Z$ exchange. 
In addition the $t$-channel contributes in selectron production via neutralinos
and in electron-sneutrino production via charginos.
The decays $\sl^- \to \ell^- \nt_i$
and  $\snu_\ell \to \ell^- \ch^+_i$  
allow a clean identification and accurate measurements of   
the primary and secondary sparticle masses and other slepton properties like   
spin, branching ratios, couplings and mixing parameters,  
see {\em e.g.} reports at LCWS~99~\cite{lcws99}.  

The masses of the first and second generation of sleptons can be determined 
within a few per mil from the isotropic two-body decay kinematics
and to the order of $100~\MeV$ 
from cross section measurements, $\sigma_{\sl \sl}\sim \beta^3$, at threshold. 

It has been suggested that $e^-e^-$ collisions provide  
a clean environment to study selectron production~\cite{feng2}. 
The main interest lies in mass determinations through threshold scans.
Selectrons associated to the same fermion helicity, 
$e^- e^- \to \se^-_R\se^-_R,  \ \se^-_L\se^-_L$, 
have a large cross section rising as $\sigma \sim\beta$,
in contrast to $e^+e^-$ annihilation. 
A simulation using the NLC machine parameters 
was presented by Heusch~\cite{heusch} and
shows, however, that this apparent advantage is depleted by initial state 
radiation and beamstrahlung effects.
The excitation curve is severely degraded and looks 
in shape (flattening of the steep rise) and magnitude very similar to the 
$e^+e^-\to \se^+\se^-$ case.
Given the considerably lower luminosity, factor of $\sim 1/5$,
it is questionable whether a competitive or even more precise
mass measurement will be achievable in comparable running times.

Studies of the third slepton generation, $\stau$ and $\snu_\tau$,
have been presented by Mizukoshi~\cite{mizukoshi}.
Due to large Yukawa couplings the stau physical eigenstates are mixed,
$\stau_1^{} = \stau_L^{} \cth_{\stau} + \stau_R^{}\sin\theta_{\stau}$ and
$\stau_2^{} = \stau_R^{} \cth_{\stau} - \stau_L^{}\sin\theta_{\stau}$,
and are no longer degenerate with the selectron and smuon masses.
While identification via decays $\stau_1\to\tau\nt_1$ will be easy and 
efficient, the background is large ($W^+W^-$ and other SUSY production)
and a mass determination using the spectra of $\tau$
decays is much less accurate. 
The mixing angle $\theta_{\stau}$ can be accessed through $\tau$ 
polarisation $\cP_\tau = \sin\theta^2_{\stau} - \cth^2_{\stau}$ 
which is measurable via the distinct energy spectra of the decays 
$\tau \to \pi\nu, \  \rho\nu$.
An accuracy of $10\%$ with $\cL=50~\fbi$ can be achieved.
Such a polarisation study has been applied to decays $\stau \to \tau \tilde G$ 
in a GMSB model, where the gravitino $ \tilde G$ is the lightest sparticle.
It allows to test the coupling $\tau\stau\tilde G$ and to set limits on
the stau lifetime in this model.

Recent results from Super-Kamiokande~\cite{superk}
suggest that neutrinos oscillate ($\nu_\mu - \nu_\tau$) and thus are massive. 
As a consequence there should also exist the 
superpartners $\snu_R$ of right-handed neutrinos (RHN). 
The addition of a new singlet neutrino field would change the predictions
for slepton masses~\cite{mizukoshi}.
RHN effects may be observable as deviations from mass relations, 
in particular those involving the third generation, {\em e.g.} 
$2\,(m^2_{\snu_R}-m^2_{\snu_\tau}) \approx m^2_{\ser} -m^2_{\stau_1}$
(up to higher order corrections). 
In order to become observable, present data require a sensitivity on mass 
measurements of 2.5\%. 
The limitation comes from the third slepton generation.
In a case study $m_{\stau_1}$ could be determined to 1.5\%
using hadronic $\tau$ decay spectra, while the accuracy for 
$m_{\snu_\tau}$ obtained from cross section measurements 
well above threshold is much worse and insufficient and needs to be improved.

\section{Charginos and Neutralinos}
Charginos and neutralinos are produced in pairs 
\begin{eqnarray*}   
  e^+e^- & \to & \ti\chi^+_i \ti\chi^-_j \qquad\qquad [i,j = 1,2]  \\
  e^+e^- & \to & \ti\chi^0_i \ti\chi^0_j \qquad\qquad \ \, [i,j = 1, \ldots,4] 
\end{eqnarray*}
via $s$-channel $\gamma / Z$ exchange and $t$-channel selectron or sneutrino
exchange. 
They are easy to detect via their decays into lighter charginos/neutralinos
and gauge or Higgs bosons or into sfermion-fermion pairs.
If these two-body decays are kinematically not possible, 
typically for the lighter chargino and neutralino,
they decay via virtual gauge bosons and sfermions,
{\em e.g.}  $\ti\chi^+_1 \to f \bar{f}' \ti\chi^0_1$ or 
$\ti\chi^0_2 \to f \bar{f} \ti\chi^0_1$.
In $R$-parity conserving MSSM scenarios the lightest neutralino $\ti\chi^0_1$ 
is stable.
Typical mass resolutions (see {\em e.g.} contributions to \cite{lcws99})
for the lighter chargino and neutralinos are expected
to be at the per mil level from decays into electrons, muons or quark jets
and below $100~\MeV$ from threshold scans, 
where the cross section rises as $\sigma_{\ch\ch} \sim \beta$.
Charged and neutral $\ch$'s are abundantly produced in decay chains
of heavy SUSY particles. By exploiting all di-lepton and di-jet mass spectra
one will be able to measure mass difference of cascade decays,
{\em e.g.} $\Delta m (\nt_2 - \nt_1)$ and $\Delta m (\ch^\pm_1 - \nt_1)$,
with a resolution of better than $50~\MeV$, essentially given
by the detector performance.

For large $\tan\beta$ the decay pattern may be very different and
the mass splitting of the $\stau$ sector may lead to a 
situation where $m_{\stau_1} < m_{\ch^\pm_1},\, m_{\nt_2}$.
Consequently the decays $\ch^+_1\to \stau^+_1\nu$
and $\nt_2 \to \stau^+_1 \tau^-$
dominate over all other decay modes via lepton or quark pairs. 
A simulation of 
$e^+e^-\to \ch^+_1\ch^-_1 \to \stau_1^+\nu\ \stau_1^-\nu 
         \to \tau^+\nu\nt_1 \ \tau^-\nu\nt_1$ 
with $m_{\ch^\pm_1}=172.5~\GeV$, $m_{\stau_1}=152.7~\GeV$,
$m_{\nt_1}=86.8~\GeV$ 
and $\tan\beta=50$ was reported by Kamon~\cite{kamon}.
Fitting the energy distribution of hadronic $\tau$ decays,
which depend on the masses of all three sparticles involved, 
results in resolutions of about $4\%$ for the $\ch^\pm_1$ and $\stau_1$ masses.
Note that cross section measurements 
are less affected by $\tau$ topologies 
and become more important for precise mass determinations in large 
$\tan\beta$ scenarios.

The properties of the chargino and neutralino systems and the extraction of 
fundamental SUSY parameters in a model independent way have been discussed by 
Kalinowski~\cite{kalinowski} and Bl\"ochinger~\cite{bloechinger}.
In the MSSM the chargino system depends on the parameters $M_2$, $\mu$ 
and $\tan\beta$. 
Charginos are composed of Winos and Higgsinos.
An easy way to access the Wino component is via $t$-channel $\snu_e$ exchange,
which couples only to left-handed electrons.
Thus the mixing parameters of the chargino system as well as the mass of the 
exchanged sneutrino can be determined by varying the beam polarisation.
If the collider energy is sufficient to produce all chargino states 
the SUSY parameters can be extracted from the masses and polarised production 
cross sections in an unambiguous way.
The neutralino system,
which is a mixture of Bino, Wino and two Higgsino fields,
depends in addition to $M_2$, $\mu$ and $\tan\beta$ 
on the $U(1)$ gaugino parameter $M_1$.
The diagonalisation of the $4 \times 4$ mass matrix is much more involved 
and a general analysis to extract the four fundamental SUSY parameters has not
yet been done. Therefore the neutralinos are primarily used to determine
$M_1$~\cite{bloechinger}. 
Further, if not directly accessible, the mass of the exchanged selectron
can be determined up to $800~\GeV$ with a resolution of $10~\GeV$.
It should be noted once more that the use of polarisation as well as 
exploiting spin correlations, 
for instance in the reaction $e^+e^-\to \nt_2\nt_1\to\ell^+\ell^-\nt_1\nt_1$,
is of great importance.

For final precision measurements the inclusion of higher order
electroweak radiative corrections will be important, 
as discussed by D\'{\i}az~\cite{diaz}. They may change the polarised cross
sections by up to 15\% and also have a strong influence on the scheme
dependent definitions of masses.
The expected sensitivity of the chargino and neutralino systems
to the parameters of two mSUGRA models are given in table~\ref{tab:chipar}.
\begin{table}[hbt] \centering 
  \caption{Estimated accuracy for the parameters $M_2$, $\mu$ and
    $\tan\beta$ from chargino 
    and $M_1$ from neutralino production
    for mSUGRA scenarios
    (based on ${\cal L}=500~\fbi$ for each $e^-$ polarisation)}
  \vspace*{2mm}
    \begin{tabular}{l c c c c}
      parameter & input  & fit value & input  & fit value  
      \\[1ex] \hdick  \\[-1.5ex]
 $M_2$   & 152\,GeV & $152\pm 1.8$\,GeV 
         & 150\,GeV & $150\pm 1.2$\,GeV \\
 $\mu$   & 316\,GeV & $316\pm 0.9$\,GeV 
         & 263\,GeV & $263\pm 0.7$\,GeV \\
 $\tan\beta$ & 3    & $3\pm 0.7$ \ \ \   
             & 30   &  $> 20$          \\
 $M_1$   &78.7\,GeV & $78.7\pm 0.7$\,GeV 
         &78.0\,GeV & $78.0\pm 0.4$\,GeV 
 \\[1ex]
    \end{tabular} 
 \label{tab:chipar}
\end{table}
$M_1$, $M_2$ and $\mu$ can be determined very precisely to one per cent
or better. Large $\tan\beta$ values are difficult to extract from
the $\chi$ systems, they are easier accessible in the $\stau$ sector.

In general, the parameters $M_1$, $M_2$ and $\mu$ 
can be complex, which also leads to $CP$ violation.
In fact, $M_2$ may be taken real, so that only two additional phases 
$\phi_\mu$ and $\phi_{M_1}$ remain.
The detection of $CP$ violating phases in MSSM models
has been discussed by Plehn~\cite{plehn}.
Limits from the electric dipole moments of the electron, neutron and mercury
suggest small phases of $\phi \lesssim 0.001$.
Measurements of chargino and neutralino masses and production cross sections
would only give a modest precision of $\delta\phi \simeq 0.1$.
A more promising approach is to construct directly $CP$ sensitive 
quantities, like triple products of momentum vectors. 
Consider for example the reaction $e^+e^-\to \nt_2\nt_1\to e^+e^-\nt_1\nt_1$. 
The distribution of the angle between the normal of the di-lepton plane
and the beam direction is expected to exhibit $CP$ asymmetries of typically
$0.1 - 1.5\%$, which is a challenge to the experiments.

A quite different $\ch$ (and $\sl$)
mass spectrum is predicted in so-called anomaly mediated  
SUSY breaking (AMSB)  models, 
where gaugino masses are no more universal but generated at one loop.
The reversed hierarchy of gaugino parameters $M_1 \sim 3\, M_2$
(in contrast to SUGRA with $M_1 \simeq 0.5\, M_2$) 
leads to near degeneracy of the lighter chargino
$\ch^\pm_1$ and the wino--like neutralino $\ch^0_1$ masses. 
Search strategies for $e^+e^- \to \ch^+_1 \ch^1_1$ production 
in AMSB models were discussed by Mrenna~\cite{mrenna}.
The signatures rely on the lifetime and decay modes of 
$\ch^\pm_1$, which depend almost entirely on the small mass difference
$\dmchi \equiv m_{\ch^{\pm}_{1}} - m_{\nt_{1}}$. 
If $\dmchi < 0.2~\GeV$ the chargino has a long lifetime yielding either 
a heavily ionising or a terminating track without visible decay products. 
For $ 0.2~\GeV < \dmchi \lesssim 2~\GeV$,
most typical of models with loop-dominated gaugino masses, 
the decay pion(s) will be detected, 
possibly associated to a secondary vertex.
The large $\gamma\gamma \to \pi\pi$  background may be suppressed
by requiring an additional photon.
If the pions have too low an energy to be detected, then one relies on a single
photon plus missing mass from $e^+e^-\to\gamma\ch^+_1 \ch^-_1$.
Once $\dmchi\gsim 2~\GeV$, the $\ch^\pm$ the signatures resemble the usual 
MSSM topologies.
With a luminosity of $\cL=50~\fbi$
the AMSB discovery potential extends over a large $\dmchi$ region almost
to the kinematic limit.

\section{SUSY parameters at high energy scales}
The precise mass measurements of sleptons, neutralinos and charginos
constitute an over-constrained set of observables, which allow the structure 
and parameters of the underlying SUSY theory to be determined.
The renormalisation group equations (RGE) relate the observable 
masses to the fundamental SUSY parameters at high energy scales.
Two different approaches were discussed by Blair~\cite{blair}.

A widely used strategy, for example at the LHC, 
is to assume a SUSY breaking scenario and then fit to the corresponding 
low-energy particle spectrum including experimental uncertainties.
If applying such a model dependent top-down approach to a specific
mSUGRA model one expects excellent accuracies for the parameters:
for the common scalar mass
$m_0= 100\pm0.09~\GeV$, for the common gaugino mass $m_{1/2}=200\pm0.10~\GeV$, 
$\tan\beta=3\pm0.02$ and for the trilinear coupling $A_0=0\pm6.3~\GeV$.
The magnitude of $\mu$ is obtained implicitly by the requirement
of electroweak symmetry breaking.
The weakness of such an approach is that scenario assumptions are effectively 
constraints in the fit and so one may miss alternative solutions or
new intermediate scales below the GUT scale.

\begin{figure}
  \centering  \vspace{-10mm}
  \epsfig{file=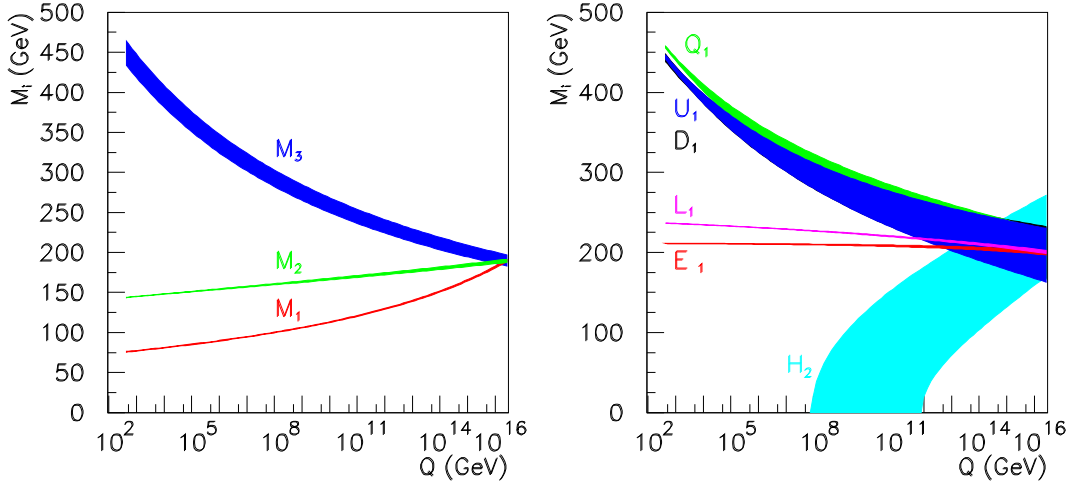,height=8cm } 
  \caption{
    Evolution of gaugino and sfermion mass parameters in a mSUGRA 
    scenario for
    $m_{0} = 200$~GeV, $m_{1/2} = 190~\GeV$, $A_{0} = 500~\GeV$, 
    $\tan\beta = 30$ and ${\rm sign}\,\mu <0 $. 
    The bands indicate 95\% CL contours.}
  \label{fig:hiE}
\end{figure}

The great advantage of an $e^+e^-$ Linear Collider is that the rich and 
precise information allows to perform a model independent analysis, 
where the structure of the theory is extrapolated from low energy 
to high energy scales via RGEs.
Input to this bottom-up approach are experimental measurements alone
without any assumption an a model.
An extrapolation of SUSY parameters from the weak scale to the GUT scale 
within a mSUGRA scenario is shown in figure~\ref{fig:hiE}. 
The gaugino mass parameters $M_{1,2,3}$ 
and the slepton mass parameters $M_{L_{1}}, M_{E_{1}}$ for the first 
and second generation are in excellent agreement with unification.
Using only LHC information would give uncertainties on the unification scale 
worse by more than an order of magnitude.
The squark parameters $M_{Q_{1}}, M_{U_{1}}, M_{D_{1}}$ and the Higgs 
parameter $M_{H_{2}}$, being less well known, still allow to test unification.
New patterns at intermediate scales would be immediately visible.

\section{Outlook}
The present Linear Collider projects JLC, NLC and TESLA will probably not 
be able to explore the full supersymmetry spectrum. A multi-TeV collider
may be necessary to complete the programme, in particular to study the 
properties of the coloured squarks and gluinos. A first look on
experimentation at CLIC has been taken by WIlson~\cite{wilson}.
A case study to search for di-leptons from SUSY processes around
$3~\TeV$ shows that edges in energy spectra from decay kinematics
are difficult to observe due to more degenerate mass
spectra as well as detector and machine effects. 
ISR and beamstrahlung effects provide a relatively wide energy spread;
but mass determinations of a couple of per cent from threshold 
scans should be feasible. 
The search strategy would probably be a bottom-up approach and slowly rise 
the cms energy above each sparticle threshold 
and make use of polarisation to improve on the signal.

\medskip
If supersymmetry is realised at low energy, an $e^+e^-$ Linear Collider 
will be an ideal instrument to explore the full portrait of the 
accessible sparticle spectrum. This workshop has shown that much progress
has been made to develop the tools 
--- experimental analysis techniques and theoretical ideas ---
to determine the sparticle properties with high accuracy. 
The LHC may discover supersymmetry and constrain its gross features. 
However, only high precision measurements at the  Linear Collider 
will be able to pin down the detailed structure of the underlying 
super\-symmetry theory.
The potential of the Linear Collider includes specifically:
\begin{itemize}
  \item[--] precise determination of sparticle masses, widths and 
    branching ratios      
  \item[--]  precise determination of couplings 
  \item[--] measurement of mixing angles in the $\tilde{t}$ 
    and  $\tilde{\tau}$  sectors 
  \item[--] determination of large $\tan\beta$ 
    in the $\tilde{\tau}$  sector 
  \item[--] determination of spin-parity $J^{PC}$ 
    and electroweak quantum numbers
  \item[--] model independent determination of SUSY parameters
\end{itemize}
It should be emphasised once more that for this ambitious programme
the highest possible luminosity is required and the 
availability of polarised beams is important.

\bigskip \bigskip \noindent 
{\it Acknowledgements} 
I want to thank all my colleagues for providing and discussing 
the material of their work presented in the many talks.
It is a pleasure to thank the organisers of LCWS~2000 for having prepared such
an excellent workshop in the stimulating atmosphere of Fermilab.

\end{document}